\documentstyle[epsfig,prd,floats,aps]{revtex}
\begin{document}
\draft

\twocolumn[\hsize\textwidth\columnwidth\hsize\csname @twocolumnfalse\endcsname

\title{Harmonic coordinate method for simulating generic singularities}
\author{David Garfinkle}
\address{
Department of Physics, Oakland University,
Rochester, Michigan 48309}
\maketitle
\null\vspace{-1.75mm}
\begin{abstract}
This paper presents both a numerical method for general relativity and an
application of that method.  The method involves the use of harmonic
coordinates in a 3+1 code to evolve the Einstein equations with scalar field
matter.  In such coordinates, the terms in Einstein's equations 
with the highest number of derivatives take a form similar to
that of the wave equation.     
The application is an exploration of the generic approach to the
singularity for this type of matter.  The preliminary results indicate
that the dynamics as one approaches the singularity is locally the dynamics
of the Kasner spacetimes.
\end{abstract}
\pacs{04.25.Dm, 04.20.Dw}

\narrowtext

\vskip2pc]
\section{Introduction}

It has long been known that harmonic coordinates are useful for
mathematical relativity.  In particular, these coordinates were
used\cite{choquet} to prove local existence of solutions of the vacuum
Einstein equation.  This usefulness of harmonic coordinates stems from their
putting Einstein's equation into a form that is similar to the 
curved spacetime wave equation.
Since many numerical techniques work well on the wave equation, one might
expect that harmonic coordinates would be used extensively in numerical 
relativity, and it is somewhat surprising that they are not.  
(However see \cite{bernd} for some recent mathematical and numerical 
work on the linearized case.  In addition, harmonic time slices have been
advocated and used in numerical relativity\cite{hs1,hs2,hs3,hs4,hs5}).
This is perhaps
due to the following drawback of harmonic coordinates: these coordinates are
solutions of the wave equation, and such solutions need not have a timelike
gradient at all points of spacetime, even if they start out with a timelike
gradient on an initial data surface.  What this means is that in harmonic
coordinates, the time coordinate will in general not remain timelike and 
this is likely to cause numerical problems.\cite{miguel,miguel2}  
As we will see later, there is
a way around this difficulty. 

One area of numerical relativity where harmonic coordinates have been used
(though somewhat unintentionally) is in the study of the approach to the
singularity, in particular in the Gowdy spacetimes.\cite{bevandvince}  
Numerical simulation
of approach to a spacetime singularity presents problems of its own.  Since
it is expected that various quantities become infinite at the singularity,
the numerical simulation will generally stop after a finite coordinate
time, the time when a surface of constant time first encounters the
singularity.  Since this first encounter generally occurs at one spatial
point, the information about the behavior of the singularity at other
spatial points will be unavailable from the numerical simulation.  The
solution to this difficulty is to choose a time coordinate that tends to
infinity as the singularity is approached.  In this way, the simulation is
not forced to end at finite coordinate time, the whole spacetime up to the
singularity is covered and the behavior of the metric as the singularity
is approached simply becomes asymptotic behavior in the limit of large
time coordinate.  For Gowdy spacetimes there is a natural choice of such
a time coordinate: these spacetimes are foliated by $T^2$s invariant under
the symmetry group.  The area of the symmetry $T^2$s goes to zero as the
singularity is approached.  Therefore minus the logarithm of this area is a
natural time coordinate that goes to infinity as the singularity is
approached.  While this method works well for the Gowdy spacetimes,
since it depends crucially on the symmetry of the Gowdy spacetimes, the
method does not seem to generalize to the case of the generic
singularity with no symmetries.  A natural generalization comes when one
notices that this time coordinate (minus the logarithm of the area of the
symmetry $T^2$s) is also harmonic.  Since in some sense one expects the
wave equation to become singular as the spacetime singularity is
approached, one might also expect a solution of the wave equation to blow
up as the singularity is approached and thus one might want to use such
a solution as the coordinate time.

What is expected to be the generic behavior of a spacetime as the
singularity is approached?  Based on studies of spacetimes with $T^2$
symmetry\cite{bevandvince,bevandme,bevandmarsha} and 
spacetimes with $U(1)$ symmetry\cite{bevandvince2}, 
the expected answer\cite{allofus} is the following: the 
singularity is expected to be spacelike and as it is approached each spatial
point is expected to ``decouple'' from the others and undergo a dynamics
corresponding to that of a homogeneous spacetime (though a different
homogeneous spacetime at each spatial point).  Which types of homogeneous
spacetime is this dynamics expected to correspond to?  For vacuum
spacetimes, it is thought that the dynamics will be oscillatory, possibly
corresponding to the Mixmaster spacetime as conjectured in \cite{bkl}.  
For many other types of matter
it is expected that as the singularity is approached the matter terms in 
the Einstein equations become negligible and the dynamics approaches that
of a vacuum spacetime.  

However for so called ``stiff matter'' ({\it i.e.} a
scalar field or a perfect fluid with equation of state $P = \rho$) it is
expected that the dynamics will not be oscillatory and will correspond to
that of a Kasner spacetime.  This expectation is greatly bolstered by a
theorem due to Andersson and Rendall\cite{larsandalan} which shows local
existence in a neighborhood of the singularity of solutions to the
Einstein-scalar equations, with the expected asymptotic behavior and with
enough degrees of freedom to be the generic solutions.  

Since the approach to the singularity is expected to be simpler for stiff
matter than for vacuum, the stiff matter case should be easiser to treat
numerically.  Therefore, in this paper we confine ourselves to a numerical 
study of the approach to the generic singularity in the Einstein-scalar
system.  Section 2 presents the equations and numerical methods used.  The
results are given in section 3.  Section 4 contains a discussion of the 
results and of other possible applications of the harmonic coordinate method.

\section{Equations and Numerical Methods}

The equations that we wish to evolve numerically are the Einstein-scalar
equations
\begin{equation}
{R_{\alpha \beta }} = 8 \pi {\nabla _\alpha}
\phi {\nabla _\beta } \phi 
\label{einsteinscalar}
\end{equation}
Here, we use the conventions of\cite{wald} including units where $c=G=1$.
As a consequence of Eq. (\ref{einsteinscalar}) and the Bianchi identities,
the scalar field must satisfy the wave equation
\begin{equation}
{\nabla _\alpha} {\nabla ^\alpha} \phi = 0
\label{wave}
\end{equation}
The Ricci tensor is given in terms of the Christoffel symbols by
\begin{equation}
{R_{\alpha \beta}} = {\partial _\gamma} {\Gamma ^\gamma _{\alpha \beta}}
 -  {\partial _\alpha} {\Gamma ^\gamma _{\gamma \beta}}  +  
{\Gamma ^\gamma _{\alpha \beta}} {\Gamma ^\nu _{\nu \gamma}}  - 
{\Gamma ^\gamma _{\alpha \nu}}{\Gamma ^\nu _{\beta \gamma}}
\label{ricdef}
\end{equation}
while the Christoffel symbols are given in terms of the metric by
\begin{equation}
{\Gamma ^\gamma _{\alpha \beta}} = 
{\textstyle {1 \over 2}}  {g^{\gamma \delta}}
\left ( {\partial _\alpha} {g_{\beta \delta}} + {\partial _\beta}
{g_{\alpha \delta}} - {\partial _\delta} {g_{\alpha \beta}}\right )
\label{chrdef}
\end{equation}

Harmonic coordinates are solutions of the wave equation.  As a generalization
of harmonic coordinates, consider coordinates that satisfy the wave equation
with source
\begin{equation}
{\nabla ^\alpha} {\nabla _\alpha } {x^\mu} = {H^\mu}
\label{genharmonic}
\end{equation}
where $H^\mu$ are specified from the beginning.  Then using
Eqs. (\ref{ricdef},\ref{chrdef},\ref{genharmonic}) we find that the Ricci
tensor is given by
\begin{eqnarray}
{R_{\alpha \beta }} = - {\textstyle {1 \over 2}} {g^{\gamma \sigma}} 
{\partial _\gamma}
{\partial _\sigma} {g_{\alpha \beta}}  +  
{\textstyle {1 \over 2}} {{C_\beta } ^{\mu
\nu}} {C_{\mu \nu \alpha }}  \cr +  
{\textstyle {1 \over 2}} {{C_\alpha} ^{\mu \nu}}
{C_{\mu \nu \beta}}  -  {\Gamma ^\gamma _{\nu \alpha }} 
{\Gamma ^\nu
_{\gamma \beta }}  -  {\partial _{(\alpha}} {H_{\beta )}}  + 
{\Gamma ^\gamma _{\alpha \beta }} {H_\gamma} 
\label{ricgenharmonic}
\end{eqnarray}
where ${C_{\alpha \mu \nu}} \equiv {\partial _\alpha} {g_{\mu \nu}}$.
Note that the second derivative terms appear only in the wave operator.  
Therefore, one might expect that Einstein's equations in this form behave
similarly to the wave equation and that numerical methods that work well
on the wave equation might work well on Einstein's equations in this form.
The reason for considering nonzero source terms $H^\mu$ in  
Eq. (\ref{genharmonic}) is that these terms allow us to change the behavior
of the time coordinate and thus may allow us to eliminate (or at least 
postpone) the behavior where the time coordinate ceases to be timelike.
In the simulations done in this paper, no source terms were needed.  
I know of no systematic way to find appropriate source terms and expect
that in the cases where they are needed, one will have to resort to
a trial and error method to find appropriate $H^\mu$.  
Note that the use of spatial harmonic coordinates could also lead to
coordinate problems.  This would occur if the gradients of the four
coordinates fail to be linearly independent.  If this sort of problem
occurs, one might expect to be able to postpone or eliminate it by
using appropriate source terms.

The numerical method used requires equations that are first order in time. 
To put the equations in such a form, we define quantities $P_{\alpha \beta}$
and $P_\phi$ given by
\begin{equation}
{P_{\alpha \beta}} = {\partial _t} {g_{\alpha \beta}}
\label{pdef}
\end{equation}
\begin{equation}
{P_\phi} = {\partial _t} \phi 
\label{pphidef}
\end{equation}
Then the Einstein-scalar equation becomes
\begin{eqnarray}
- {g^{00}} {\partial _t} {P_{\alpha \beta }} = 2 {g^{0k}} {\partial _k}
{P_{\alpha \beta }}  +  {g^{ik}} {\partial _i} {\partial _k} 
{g_{\alpha \beta }}  \cr +   16 \pi {\partial _\alpha } \phi {\partial _\beta}
\phi  +  2 {\partial _{(\alpha }} {H_{\beta )}}  -  2 
{\Gamma ^\gamma _{\alpha \beta }} {H_\gamma} 
\cr -  {{C_\alpha } 
^{\mu \nu}} {C_{\mu \nu \beta}}   -  {{C_\beta } ^{\mu \nu}}
{C_{\mu \nu \alpha}}  +  2 {\Gamma ^\gamma _{\nu \alpha }}
{\Gamma ^\nu _{\gamma \beta }}
\label{evolvep}
\end{eqnarray}
The wave equation for $\phi$ becomes
\begin{equation}
- {g^{00}} {\partial _t} {P_\phi} = 2 {g^{0k}} {\partial _k} {P_\phi}  + 
{g^{ik}} {\partial _i} {\partial _k} \phi  -  {g^{\alpha \beta }}
{\Gamma ^\gamma _{\alpha \beta }} {\partial _\gamma} \phi
\label{evolvepphi}
\end{equation}
The full set of equations that are evolved in the computer code are
Eqs. (\ref{pdef}-\ref{evolvepphi}).

We now turn to the numerical methods used to evolve these equations. 
Spatial derivatives are approximated by centered differences.  
The variables are evolved in time using a three step
iterated Crank-Nicholson
(ICN) method.\cite{matt,saul,golm} 
This works as follows: evolution equations
of the form ${\partial _t} S = W(S)$ for some set of variables $S$
are approximated as 
\begin{equation}
{S^{n+1}} = {S^n} + {{\Delta t} \over 2} \left [ W({S^n}) + W({S^{n+1}})
\right ]
\label{cn}
\end{equation}
where $S^n$
is the value of $S$ at time step $n$ 
and $\Delta t$ is the time step.
Then, using $S^n$ as an initial guess for $S^{n+1}$, Eq. (\ref{cn})
is iterated three times.

The spacetimes we consider have topology ${T^3} \times R$.  Each spatial
slice has topology $T^3$.  In terms of the spatial coordinates, this means
that $0 \le x \le 2 \pi$ with $0$ and $2 \pi$ identified (and
correspondingly for $y$ and $z$).  This topology is implemented numerically as
follows: a spatial coordinate $x$ has $N$ grid points.  The variables on
points from $2$ to $N-1$ are evolved using the evolution equations.  The
variables on point $1$ are set to the values at point $N-1$ while at point
$N$ they are set to the values at point $2$.

\section{Results}

All runs were done in double precision on Compaq XP1000 workstations and on
the NCSA Origin 2000.  The time step was $\Delta t = \Delta x/2$.
While the source terms $H^\mu$ should be helpful in keeping
the coordinates well behaved, we did not need them for the cases studied here
and all runs have ${H^\mu} = 0$.

\begin{figure}[bth]
\begin{center}
\makebox[3.0in]{\psfig{file=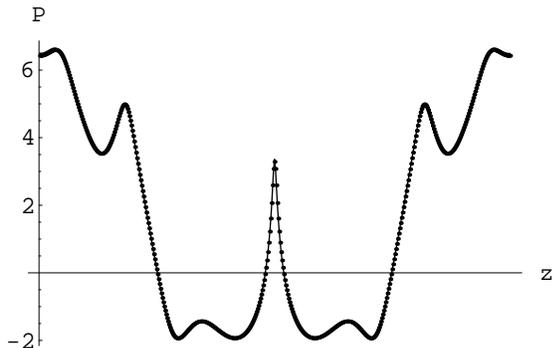,width=3.0in}}
\caption{comparison of the quantity $P$ as found using a Gowdy code
and the 3+1 harmonic code}
\end{center}
\label{eps1}
\end{figure}

Before exploring the generic singularity, we would like to test the code.
One way of doing this is to run the code on cases with symmetry where
results are already known.  Such a case is the Gowdy spacetimes.  These
have the form\cite{bevandvince}
\begin{eqnarray}
d {s^2} = {e^{(\tau - \lambda )/2}} \left [ - {e^{- 2 \tau }} d {\tau ^2}
+ d {z ^2} \right ] \cr
+ {e^{ - \tau }} \left [ {e^P} d {x^2} + 2 {e^P} Q d x d y 
+ \left ( {e^P} {Q^2} + {e^{- P}}\right ) d {y^2} \right ]
\label{gowdy}
\end{eqnarray}
Here, $P,Q$ and $\lambda$ are functions of $\tau$ and $z$. It follows from
Eq. (\ref{gowdy}) that the coordinates ($\tau ,x,y,z$) are all solutions 
of the wave equation.  Therefore, they are harmonic coordinates.  The
vacuum Einstein equations for the Gowdy spacetimes are\cite{bevandme}
\begin{equation}
{\partial _\tau} {\partial _\tau } P - {e^{- 2 \tau}} {\partial _z} 
{\partial _z} P - {e^{2P}} \left [ {{({\partial _\tau} Q)}^2} -
{e^{ - 2 \tau}} {{({\partial _z} Q)}^2}\right ] = 0
\label{gowdyp}
\end{equation}
\begin{equation}
{\partial _\tau} {\partial _\tau } Q - {e^{- 2 \tau}} {\partial _z} 
{\partial _z} Q + 2 \left ( {\partial _\tau } P {\partial _\tau } Q
- {e^{- 2 \tau}} {\partial _z} P {\partial _z} Q \right ) = 0
\label{gowdyq}
\end{equation}
plus constraint equations that determine $\lambda$ once $P$ and $Q$ are
known.  

To test the 3+1 code, we evolve Eqs. (\ref{gowdyp}) and 
(\ref{gowdyq}) using a 1+1 code.  Then we evolve the same initial data
with the 3+1 harmonic code and compare the results.  To do the comparison,
note that $P= \tau + \ln {g_{xx}}$ so that it is straightforward to 
compare the values of $P$ produced by the two codes.  
For these simulations 500 gridpoints were used in the 1+1 code. 
In the 3+1 code, 3 gridpoints were used in the $x$ direction, 3 gridpoints
in the $y$ direction and 500 gridpoints in the $z$ direction.
For the comparison
the initial data used is $P=0, \; {\partial _\tau} P = 5 \cos z , 
\; Q = \cos z ,  \; {\partial _\tau} Q = 0$.
These data are evolved until $\tau = \pi $ and the results
for the comparison are given in Fig. 1.  Here, the solid line represents
the 1+1 evolution of the Gowdy equations, while the dots represent the full
3+1 evolution using the harmonic code.  There is clearly agreement between
the two.

\begin{figure}[bth]
\begin{center}
\makebox[3.0in]{\psfig{file=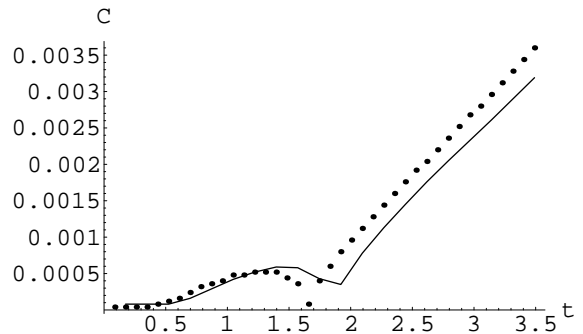,width=3.0in}}
\caption{convergence test involving the constraint}
\label{eps2}
\end{center}
\end{figure}

Before presenting the results of another code test, we turn to the question
of finding initial data without symmetries.  On an initial data
surface, the intrinsic metric $h_{ij}$ and the extrinsic curvature 
$K_{ij}$ must satisfy the constraint equations
\begin{equation}
{D_i} {{K^i}_j} - {D_j} K = 8 \pi  {\dot \phi} {D_j} \phi
\label{momentum}
\end{equation}
\begin{equation}
{^{(3)}}R + {K^2} - {K^{ij}}{K_{ij}} = 8 \pi \left [ {{\dot \phi}^2} 
+ {D^i} \phi {D_i} \phi \right ] 
\label{hamiltonian}
\end{equation}
(Note that we use the convention of\cite{wald} for the sign of $K_{ij}$ which 
is opposite to the convention usually used in numerical relativity).
Here, an overdot denotes derivative along the normal to the surface,
and $D_i$ and ${^{(3)}}R$
are respectively the covariant derivative and scalar curvature of $h_{ij}$.
Given a solution of Eqs. (\ref{momentum}) and (\ref{hamiltonian}),
we produce initial data for evolution in harmonic coordinates as follows:
For spatial directions $i$ and $j$ we have ${g_{ij}}={h_{ij}}, \; 
{g_{i0}}={g_{0i}}=0, \; {g_{00}}=-1 ,\; {P_{ij}}= 2 {K_{ij}}$.  The 
remaining components of $P_{\alpha \beta}$ are solved for using 
Eq. (\ref{genharmonic}).

We want to find a solution of Eqs. (\ref{momentum}) and
(\ref{hamiltonian}) that is simple but has no symmetries and has some 
free parameters.  We choose $\phi =0$ and $h_{ij}$ equal to the flat
Euclidean metric in the usual coordinates.  For the extrinsic curvature
we choose
\begin{eqnarray}
{K_{xx}} =  \left ( {b_1} + {a_2} \cos y + {a_3} \cos z
\right ) /2
\cr
{K_{yy}} =  \left ( {b_2} + {a_1} \cos x - {a_3} \cos z
\right ) /2
\cr
{K_{zz}} =  \left ( {b_3} - {a_1} \cos x - {a_2} \cos
y \right ) /2
\cr
{K_{xy}}={K_{yx}} = \left ( {c_1} \cos z \right ) /2
\cr
{K_{xz}}={K_{zx}} = \left ( {c_2} \cos y \right ) /2
\cr
{K_{yz}}={K_{zy}} = \left ( {c_3} \cos x \right ) /2
\label{kdef}
\end{eqnarray}
Here, the quantities ${a_i},\; {b_i}$ and $c_i$ are constants that are
free parameters.  It is straightforward to show that the extrinsic
curvature of Eq. (\ref{kdef}) with our choice of initial $h_{ij}$
and $\phi$ satisfies Eq. (\ref{momentum}).  Equation (\ref{hamiltonian})
then becomes an agebraic equation for $P_\phi$ which can be solved provided
that the left hand side of the equation is positive.  Note that if all the
$a_i$ and $c_i$ are zero and all the $b_i$ are equal, then the initial data
evolve to the spatially flat Robertson-Walker spacetime with scalar field
matter.  Thus, this family of data can be thought of as Robertson-Walker
with large gravitational and scalar waves.

We now consider a convergence test involving a constraint that comes
from the use of harmonic coordinates.  
Define the quantities $C^\mu$ by
\begin{equation}
{C^\mu} = {g^{\alpha \beta}} {\Gamma ^\mu _{\alpha \beta}} 
\label{cnstr}
\end{equation}
(this is the appropriate form of the constraint for the case where the
source term $H^\mu$ vanishes.  For the general case, the constraint would
be given by 
${C^\mu} = {g^{\alpha \beta}} {\Gamma ^\mu _{\alpha \beta}} + {H^\mu}$).
Then from Eq. (\ref{genharmonic}) it follows that ${C^\mu}=0$.
Since we are solving the evolution equations by approximating them by
finite difference equations, the quantities $C^\mu$ as evaluated
by the computer code will not be zero because of errors due to the
finite grid spacing $\Delta x$.  
Define
the quantity $C$ by
\begin{equation}
C = {{\left |{{\int {\sqrt g} {g_{\alpha \beta }} {C^\alpha}
{C^\beta} dx dy dz} \over {\int {\sqrt g} dx dy dz}} \right | }^{1/2}}
\end{equation}
This quantity is a type of measure of the average size of the constraint. 
Figure 2 shows a plot of  $C$ {\it vs.} time.  
The parameters used are ${a_i}=(0.1,0.1,0.2), \; {b_i}=(0,-0.5,-0.5)$
and ${c_i}=(0,0,0)$.
Here the curve
corresponds to a run done with 20 gridpoints in each
spatial direction, while the dots correspond to a run with 
38 gridpoints in each spatial direction (which gives half the  
grid spacing) and with
$C$ multiplied by 4.  The results show second order convergence.

\begin{figure}[bth]
\begin{center}
\makebox[3.0in]{\psfig{file=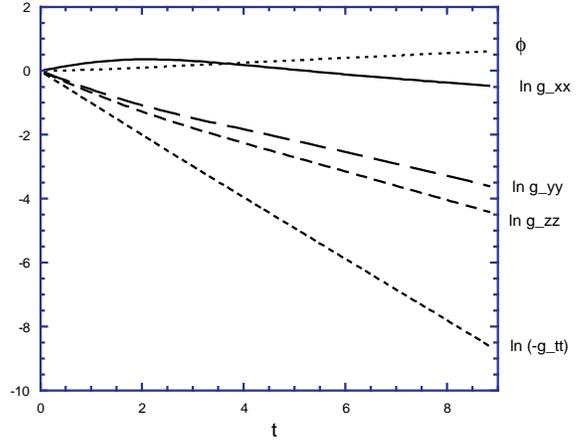,width=3.0in}}
\caption{behavior of metric components and scalar field as the
singularity is approached at the spatial point ($0,0,0$)}
\label{eps3}
\end{center}
\end{figure}

\begin{figure}[bth]
\begin{center}
\makebox[3.0in]{\psfig{file=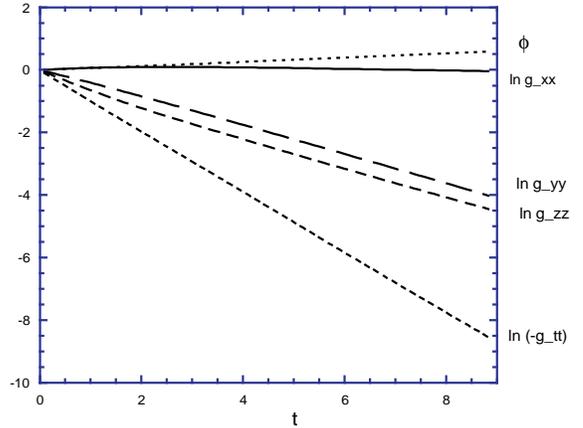,width=3.0in}}
\caption{behavior of metric components and scalar field as the
singularity is approached at the spatial point ($0,\pi /4,\pi /2$)}
\label{eps4}
\end{center}
\end{figure}

We now consider the approach to the singularity.  To see what is expected,
it is helpful to consider the Kasner spacetime in harmonic coordinates.
This is given by
\begin{equation}
d {s^2} = - {e^{({q_1}+{q_2}+{q_3})\tau}} \; d {\tau ^2}
+ {e^{{q_1} \tau }} d {x^2} + {e^{{q_2} \tau }} d {y^2}
+ {e^{{q_3} \tau }} d {z^2}
\end{equation}
where the $q_i$ are constants
This metric is generally a solution of the Einstein-scalar equations; but
for the case where $ {\left ( \sum {q_i} \right ) }^2 = \sum {q_i ^2} $
it is a vacuum spacetime.  (In the usual treatment of 
vacuum Kasner spacetimes, one defines the quantities 
${p_i}={q_i}/(\sum {q_m})$ and then obtains the condition
$ \sum {p_i} = \sum {p_i ^2} = 1$).
Note that in the vacuum case one cannot have
all three directions contracting but that this is possible for the 
Einstein-scalar case.  Note also that the metric components are
exponential functions of time.  It also turns out that the scalar field
is a linear function of time.  Thus, if the behavior of a generic solution near
the singularity is local Kasner, then we should expect  metric components that
are exponential functions of time, with the exponent depending on space.  We
should also expect a scalar field that is a linear function of time with the
slope depending on space.

Results on the approach to the singularity are shown in Figs. 3 and 4. 
The run was done with 34 grid points in each spatial directions.  The
parameters ${a_i},\; {b_i}$ and $c_i$ are the same as for the convergence
test.
Here the scalar field and the logarithms of the diagonal metric components
are plotted as functions of time.  Figure 3 corresponds to the spatial point
$(x,y,z)=(0,0,0)$ while for Fig. 4 the point is $(0,\pi /4,\pi /2)$.    
Note that as the singularity is approached
these quantities all become linear functions of time. The differences 
between Figs. 3 and 4 show that there is a spatial dependence of the
approach to the singularity.  Note from Fig. 3 that though the $x$
direction is initially expanding, eventually all three directions contract.
This is what one would expect if the metrics of\cite{larsandalan} 
represent the generic behavior near the singularity since these metrics
have all three directions contracting in a neighborhood of the singularity.  

\section{Discussion} 

While this study is somewhat preliminary, it indicates that harmonic
coordinates can be a useful tool in numerical relativity.  Though, in
principle coordinate problems could occur, this did not happen in the
cases studied here, even though they involved very strong fields. 
Furthermore, the use of the source terms $H^\mu$ may cure such problems
if they arise.

As for the behavior of generic singularities, the numerical
results indicate that solutions of the form proved in
reference\cite{larsandalan} to exist in a neighborhood of the singularity
also exist globally.  Thus such solutions are likely to describe the
generic singularity in the stiff matter case.

There are several projects for which the methods of this paper could
be used.  One is to do a more extensive study of the singularity in
the Einstein-scalar case, with a more thorough exploration of the
evolution corresponding to various values of the parameters in the initial
data of the previous section.  It would also be helpful to evolve for a
longer time.

Another project is to remove the scalar field and study the approach to
the singularity of the generic vacuum spacetime.  Here, the behavior is
expected to be more complicated and a treatment will probably require
more spatial resolution to resolve the expected sharp features, as well
as longer evolution in time to see the expected oscillatory behavior.

Yet another project is to study the behavior of asymptotically flat
spacetimes rather than closed cosmologies.  Here, the closed cosmologies
were studied partly for simplicity.  The periodic boundary conditions are
simple to implement and completely consistent with Einstein's equations.
In contrast, for an outer boundary at a finite distance in an asymptotically
flat spacetime, one needs to put some sort of outgoing wave boundary
condition.  Such conditions are usually not consistent with Einstein's
equation (it is known how to have a consistent boundary condition\cite{nagy}
but this condition is quite complicated).  These inconsistent conditions
may lead to numerical instability.  Since harmonic coordinates make
Einstein's equation look like the wave equation, simple outgoing wave
boundary conditions that work numerically with the wave equation might
be expected to ``work'' (at least in the sense of not causing numerical 
instability) for Einstein's equation.

Since much work has been done on numerical simulations of asymptotically
flat spacetimes using the standard Arnowitt-Deser-Misner 
(ADM)\cite{adm} approach, it is helpful to make
comparisons with this approach.  In the ADM approach, the spacetime
metric is written as
\begin{equation}
d {s^2} = - {\alpha ^2} d {t^2} + {h_{ij}} ( d {x^i} + {\beta ^i} dt)
(d {x^j} + {\beta ^j} dt)
\label{admmetric}
\end{equation}
Einstein's equations are written as an evolution equation for the spatial
metric $h_{ij}$ (and the extrinsic curvature $K_{ij}$) while the gauge
choice results in equations for the lapse $\alpha$ and the shift $\beta ^i$.
This framework is sufficiently general to accomodate the use of harmonic
coordinates, which correspond to the following equations for lapse and
shift
\begin{equation}
{\partial _t} \alpha - {\beta ^i} {\partial _i} \alpha =  K {\alpha ^2}
\label{harmoniclapse}
\end{equation}
\begin{equation}
{\partial _t} {\beta ^i} - {\beta ^m} {\partial _m} {\beta ^i} = 
{h^{mn}} {{{\bar \Gamma}^i}_{mn}} {\alpha ^2}
\label{harmonicshift}
\end{equation}
(Note that the sign of $K$ in equation (\ref{harmoniclapse}) comes from 
the convention of \cite{wald}).
Here ${{\bar \Gamma}^i}_{mn}$ is the Christoffel symbol associated with
$h_{ij}$.  (Equations (\ref{harmoniclapse}) and (\ref{harmonicshift}) 
hold for the case of vanishing source term $H^\mu$.  Similar equations
hold in the case of nonzero $H^\mu$).  The use of equations 
(\ref{harmoniclapse}) and (\ref{harmonicshift}) in an ADM code is not
precisely equivalent to the approach of this paper.  The reason for this
is that equations (\ref{harmoniclapse}) and (\ref{harmonicshift}) 
directly solve the constraint 
${g^{\alpha \beta}} {\Gamma ^\mu _{\alpha \beta}} = 0 $, while in our 
approach, this constraint is used to change the form of the evolution
equations.  Nonetheless, it would be interesting to use 
equations (\ref{harmoniclapse}) and (\ref{harmonicshift}) 
in an ADM code to see how
that compares with other choices of lapse and shift.  In particular, one
might expect better stability properties and compatibility with a simple
outgoing wave boundary condition.  (Though note that such improvements
can also be obtained using the BSSN approach\cite{bssn1,bssn2,bssn3}).   

Another desired feature of a numerical code is the ability to treat 
black holes, and this sometimes requires black hole excision.  While
harmonic coordinates are singularity avoiding\cite{hs1} they are just
barely so and come arbitrarily close to a singularity.  Thus the need
for excision in an approach that uses harmonic coordinates should be
at least as great as in the standard approach.  Nonetheless, one might
hope that excision itself would be easier to implement using the approach 
of this paper.  This is because no elliptic equations are involved 
and the light cone of the wave operator is the same as that of the
physical metric. 

All these projects are work in progress and preliminary results from them
are promising.  Thus, I expect that harmonic coordinates will become a useful
tool in numerical relativity.

\section{Acknowledgements}

I would like to thank Helmut Friedrich, Beverly Berger, Matt Choptuik,
G. Comer Duncan, Lars Andersson and Alan Rendall 
for helpful discussions.  I would also like to thank the Albert Einstein
Institute for hospitality.  This work
was partially supported by NSF grant PHY-9988790 to Oakland University. 
Some of the computations were performed at the National Center for 
Supercomputing Applications (Illinois).

\end{document}